\documentclass[12pt,a4paper]{article}
\usepackage{graphicx}
\usepackage{float} %
\usepackage{subfig}
\usepackage{amssymb}
\usepackage{amsmath}
\newcommand\rf[1]{(\ref{eq:#1})}
\newcommand\lab[1]{\label{eq:#1}}
\newcommand\nonu{\nonumber}
\newcommand\br{\begin{eqnarray}}
\newcommand\er{\end{eqnarray}}
\newcommand\be{\begin{equation}}
\newcommand\ee{\end{equation}}

\newcommand\lb{\lbrack}
\newcommand\rb{\rbrack}

\renewcommand\({\left(}
\renewcommand\){\right)}
\renewcommand\v{\vert}                     

\newcommand\bc{\begin{center}}
\newcommand\ec{\end{center}}

\renewcommand\a{\alpha}
\renewcommand\b{\beta}

\newcommand\vareps{\varepsilon}

\newcommand\G{\Gamma}

\newcommand\h{\frac{1}{2}}
\renewcommand\k{\kappa}
\renewcommand\l{\lambda}

\newcommand\m{\mu}
\newcommand\n{\nu}

\newcommand\vp{\varphi}

\newcommand\pa{\partial}

\newcommand\pr{\prime}

\renewcommand\th{\theta}

\newcommand\cM{{\mathcal M}}

\newcommand{\ct}[1]{\cite{#1}}
\newcommand{\bib}[1]{\bibitem{#1}}

\newcommand\PRL[3]{\textsl{Phys. Rev. Lett.} #2; #1: #3}
\newcommand\NPB[3]{\textsl{Nucl. Phys.} #2; B#1; #3}

\newcommand\PRD[3]{\textsl{Phys. Rev.} #2; D#1; #3}

\newcommand\PLB[3]{\textsl{Phys. Lett.} #2; #1B: #3}
\newcommand\CQG[3]{\textsl{Class. Quantum Grav.} #2; #1: #3}

\newcommand\AoP[3]{\textsl{Ann. of Phys.} #2; #1: #3}

\newcommand\IJMPA[3]{\textsl{Int. J. Mod. Phys.} #2; A#1: #3}

\newcommand\MPLA[3]{\textsl{Mod. Phys. Lett.} #2; A#1: #3}

\title{Fully Explorable Horned Particles Hiding Charge}
\author{E. I. Guendelman\footnote{Email:guendel@bgu.ac.il} \space\ and M. Vasihoun\footnote{Email:mahary@bgu.ac.il}\\ \\
\emph{Physics Department, Ben Gurion University, Beer Sheva, Israel}}

\begin{document}

\maketitle

\begin{abstract}
The charge-hiding effect by a horned particle, which was studied for the case where gravity/gauge-field system is self-consistently interacting with a charged lightlike brane (LLB) as a matter source, is now studied for the case of a time like brane. From the demand that no surfaces of infinite coordinate time redshift (horizons) appear in the problem we are lead now to a completly explorable horned particle space for traveller that goes through the horned particle (as was the case for the LLB) but now also in addition to this, the horned region is fully visible to a static external observer. This requires negative surface energy density for the shell sitting at the throat. We study a gauge field subsystem which is of a special non-linear form containing a square-root of the Maxwell term and which previously has been  shown to produce a QCD-like confining gauge field dynamics in flat space-time. The condition of finite energy of the system or asymptotic flatness on one side of the horned particle implies that the charged object sitting at the throat expels all the flux it produces into  the other side of the horned particle, which turns out to be of a ``tube-like'' nature. An outside  observer in the asymptotically flat universe detects, therefore, apparently neutral object. The hiding of the electric flux behind the tube-like region of a horned particle is the only possible way that a truly charged particle can still be of finite energy, in a theory that in flat space describes confinement. This points to the physical relevance of such solutions, even though there is the need of negative energy density at the throat of the horned particle, which can be of quantum mechanical origin.
\end{abstract}





\section{Introduction}
\label{intro}
The charge-hiding effect by a horned particle, which are spaces containing a space where asymptotically $r \rightarrow \infty $ connected to a space  where $r=constant$ and this is called  a "horned" region because there is another cordinate there (that replaces $r$ which is constant and cannot be used as a coordinate therefore) which runs along an infinite range\footnote{In \ct{hidingLLB} we called these spaces wormholes, but it has been pointed out to us that they are more correctly described as "horned particles" as done for the spacetimes studied in \ct{horned particle-1}, \ct{horned particle-2} that have large "tube" like structre also. Somewhat similar are the so called "Gravitational bags", where some extra dimensions grows very large at the center of the four dimensional projected metric \ct{GB-1}, \ct{GB-2},\ct{GB-3}}, which was studied for the case where gravity/gauge-field system is self-consistently interacting with a charged lightlike brane (LLB) as a matter source \ct{hidingLLB}, is now studied for the case of a  time like brane, where we demand that no surfaces of infinite coordinate time redshift\footnote{Such surface of infinite redshift is in fact a horizon at $r=r_h$, so the resulting object is similar to a black hole, but not exactly, in our previous studies of wormholes constructed this way, because there is no "interior", that is $r>r_h$ everywhere in these previous studies and since $r>r_h$ everywhere, there is no possibility of "collapse" to $r=0$ } appear in the problem, leading therefore to a completely explorable horned particle space, according to not only the traveller that goes through the horned particle space (as was the case for the LLB), but also to a static external observer. This requires negative surface energy density for the shell sitting at the throat of the horned particle. We study a gauge field subsystem which is of a special non-linear form containing a square-root of the Maxwell term and which previously has been  shown to produce a QCD-like confining gauge field dynamics in flat space-time. The condition of finite energy of the system or asymptotic flatness on one side of the horned particle implies that the charged object  sitting at the ``throat'' expels all the flux it produces into the other side of the horned particle, which turns out to be of a ``tube-like'' nature. An outside  observer in the asymptotically flat universe detects, therefore, a neutral object. The hiding of the electric flux behind the horned region of the particle appears to be the only possible way that a truly charged particle can still be of finite energy, which points to the physical relevance of such solutions, even though there is the need of negative energy density at the throat, which can be of quantum mechanical origin.

This effect is indeed the opposite to the famous Misner-Wheeler ``charge without charge'' effect \ct{misner-wheeler}, one of the most interesting physical phenomena produced by wormholes.
Misner and Wheeler realized that wormholes connecting two asymptotically flat 
space times provide the possibility of existence of electromagnetically
non-trivial solutions, where the lines of force of the electric field flow from one 
universe to the other without a source and giving the impression of being 
positively charged in one universe and negatively charged in the other universe.
Wormholes may be classified according to their  traversability properties, which can be addressed according to whether a "traveller" that attempts to cross from one side of the wormhole throat 
to the other to the other side can do so in a finite proper time (the travellers proper time), or one may require this, but in addition that a static observer, which uses the coordinate time, will see the traveller go from one side of the throat and come back in a finite coordinate time.   For a detailed account of the general theory of traversable wormholes according to this second, most stringent definition, we refer to Visser's book \ct{visser-book} (see also \ct{visser-1,visser-2} and some more recent accounts \ct{WH-rev-1}--\ct{bronnikov-2}.

In contrast to the Misner and Wheeler effect using wormhole, the charge-hiding effect by a horned particle, a genuinely charged matter source of gravity and electromagnetism may appear {\em electrically neutral} to an external observer. In previous publications it  has been shown that this phenomenon  takes place in a gravity/gauge-field system self-consistently coupled to a charged lightlike brane as a matter source, where the gauge field subsystem is of a special non-linear form containing a square-root of the Maxwell term \ct{hidingLLB}. The latter has been previously shown \ct{GG-1}--\ct{GG-6} to produce a QCD-like confining (``Cornell'' \ct{cornell-potential-1}--\ct{cornell-potential-3}) potential in flat space-time. There the lightlike brane, at the ``throat'' connects a ``universe'' (where $r\rightarrow\infty$) with a ``universe'' (where $r=Const$), is electrically charged, however all of its flux flows into the "tube-like universe'' only. No Coulomb field is produced in the ``universe'' containing the $r\rightarrow\infty$ region, therefore, the horned particle hides the charge from an external observer in the latter ``universe''.

In the case where lightlike branes are present as source of the system, sitting at the throat, there is at the throat a surface of infinite redshift for the coordinate time, so the horn is accessible only according to the weaker definition that the "traveller" that attempts to cross from one side of the horn throat to the other side can do so in a finite proper time. If we want to assure the more stringent definition of traversability or accessibility, we must use time like branes, this will be done here, in the framework of the thin-wall approach to domain walls or  thin shells coupled to gravity \ct{Israel-66}.

The gravity/gauge-field system with a square-root of the Maxwell term 
was recently studied in \ct{grav-cornell} (see the brief review in Section 2 below) where the following interesting new features of the pertinent static spherically symmetric solutions have been found:\\
(i) appearance of a constant radial electric field (in addition to the Coulomb one)
in charged black holes within Reissner-Nordstr{\"o}m-de-Sitter-type 
and/or Reissner-Nordstr{\"o}m-{\em anti}-de-Sitter-type 
space-times, in particular, in electrically neutral black holes with 
Schwarzschild-de-Sitter 
and/or Schwarzschild-{\em anti}-de-Sitter 
geometry;\\
(ii) novel mechanism of {\em dynamical generation} of cosmological constant
through the nonlinear gauge field dynamics of the ``square-root'' Maxwell term;\\
(iii) appearance of confining-type effective potential in charged test particle 
dynamics in the above black hole backgrounds.

In Section 3 of the present paper we review  the results of \ct{hidingLLB} concerning tube or  Levi-Civita-Bertotti-Robinson \ct{LC-BR-1}--\ct{LC-BR-3}, \textsl{i.e.} type solutions, with space-time geometry of the form $\cM_2 \times S^2$ with $\cM_2$ being a two-dimensional anti-de Sitter, Minkowski or de Sitter space depending on the relative strength of the electric field w.r.t. coupling of the square-root Maxwell term. 

In previous papers \ct{LL-main-1}--\ct{beograd-2010} an explicit reparametrization invariant world-volume Lagrangian formulation of lightlike $p$-branes was used to construct various types of wormhole, regular black hole and lightlike braneworld solutions in $D\!=\!4$ or higher-dimensional asymptotically flat or asymptotically anti-de Sitter bulk space-times. In particular, in refs.\ct{BR-kink}--\ct{beograd-2010} it has been shown that lightlike branes can trigger a series of transitions of space-time regions, \textsl{e.g.}, from ``tube-like'' Levi-Civita-Bertotti-Robinson spaces to non-compact Reissner-Nordstr{\"o}m or Reissner-Nordstr{\"o}m-de-Sitter region or {\sl vice versa}. Let us note that wormholes with ``tube-like'' structure (and regular black holes with ``tube-like'' core) have been previously obtained  in 
refs.\ct{eduardo-wh}--\ct{zaslavskii-3}. 


Although light like branes will not be used in this paper, one should point out nevertheless the essential role of the proper world-volume Lagrangian formulation of
lightlike branes which manifests itself most clearly in the correct self-consistent 
construction \ct{LL-main-5,Kerr-rot-WH-2} of the simplest wormhole solution first 
proposed by Einstein and Rosen \ct{einstein-rosen} -- the Einstein-Rosen ``bridge'' wormhole.
Namely, in refs.\ct{LL-main-5,Kerr-rot-WH-2} it
has been shown that the Einstein-Rosen ``bridge'' in its original formulation
\ct{einstein-rosen} naturally arises as the simplest particular case of {\em static} 
spherically symmetric wormhole solutions produced by lightlike branes as
gravitational sources, where the two identical ``universes'' with Schwarzschild
outer-region geometry are self-consistently glued together by a lightlike brane occupying
their common horizon -- the wormhole ``throat''. An understanding of this
picture within the framework of Kruskal-Szekeres manifold was subsequently
provided in ref.\ct{poplawski}, which involves Rindler's elliptic
identification of the two antipodal future event horizons. The system resembles black hole in the sense that there is a surface of infinite redshift at $r=r_h$ but unlike the standard black hole there is no $r<r_h$, only two regions glued and for both $r>r_h$.

At this point let us strongly emphasize that the original notion of 
``Einstein-Rosen bridge'' in ref.\ct{einstein-rosen} is qualitatively different from 
the notion of ``Einstein-Rosen bridge'' defined in several popular textbooks ({\sl e.g.}, 
refs.\ct{MTW,Carroll}) using the Kruskal-Szekeres manifold, where the ``bridge'' has 
{\em dynamic} space-time geometry. Namely, the two regions in 
Kruskal-Szekeres space-time corresponding to the two copies of outer Schwarzschild 
space-time region ($r>2m$) (the building blocks of the original static Einstein-Rosen 
``bridge'') and labeled $(I)$ and $(III)$ in ref.\ct{MTW} are generally
{\em disconnected} and share only a two-sphere (the angular part) as a common border
($U=0, V=0$ in Kruskal-Szekeres coordinates), whereas in the original Einstein-Rosen
``bridge'' construction \ct{einstein-rosen} the boundary between the two identical 
copies of the outer Schwarzschild space-time region ($r>2m$) is a three-dimensional 
lightlike hypersurface ($r=2m)$. 

In Section 4 below we consider the matching of an external solution, containing $r\rightarrow\infty$ region, to a tube like solution, where $r=const$, through a time like brane which will serve as a matter source of gravity and (nonlinear) electromagnetism. Then in section 5 we recognize the interesting phenomenon that for asymptotic flatness ( and therefore for the configuration to be recognized as a legitimate finite energy excitation from flat space) the charged particle has to redirect all the flux it produces in the direction of the tube region. This new charge ``confining'' phenomena is entirely due to the presence of the ``square-root'' Maxwell term, which assigns infinite energy to any configuration that does not hide the flux (i.e. that does not send all the electric flux in the direction of the tube region).

\section{Lagrangian Formulation. Spherically Symmetric Solutions}
\label{lagrange}
In Refs. \ct{GG-1}--\ct{GG-6} a flat space-time model of nonlinear gauge field system
with a square-root of the Maxwell term was considered ($f$ below is a positive constant that sets the scale for confinement effects in the theory)
\br
S = \int d^4 x L(F^2) \quad ,\quad
L(F^2) = - \frac{1}{4} F^2 - \frac{f}{2} \sqrt{- F^2} \; \lab{flatmodel}
\\
F^2 \equiv F_{\m\n} F^{\m\n} \quad ,\quad 
F_{\m\n} = \pa_\m A_\n - \pa_\n A_\m \; .
\nonu
\er

The equations of motion are
\be
\pa_\n \( \( \sqrt{-F_{\a\b}F^{\a\b}}-f\) \frac{F^{\m\n}}{\sqrt{-F_{\a\b}F^{\a\b}}} \)=0 
\lab{flat GG-eqs}
\ee       
Then, assuming spherical symmetry and time independence, we find that, in addition to a Coulomb like piece, a linear term proportional to $f$ is obtained for $A^{0}$, which is of the form of the well-known “Cornell” potential \ct{cornell-potential-1}--\ct{cornell-potential-3} in quantum chromodynamics (QCD). Furthermore, these equations are consistent with the 't Hooft criteria for perturbative confinement. In fact, in the infrared region the above equation implies that
\be
F^{\m\n}=f\frac{F^{\m\n}}{\sqrt{-F_{\a\b}F^{\a\b}}}
\ee
plus negligible terms in the infrared sector. Interestingly enough, for a static source, this automatically implies that
the chromoelectric field has a fixed amplitude. Confinement is obvious then, since in the presence of two external
oppositely charged sources, by symmetry arguments, one can see that such a constant amplitude chromoelectric
field must be in the direction of the line joining the two charges. The potential that gives rise to this kind of field
configuration is of course a linear potential. Also, for static field configurations the model \rf{flatmodel} yields the following electric displacement field $\overrightarrow{D} = \overrightarrow{E} - \dfrac{f}{\sqrt{2}}\dfrac{\overrightarrow{E}}{|\overrightarrow{E}|}$. The pertinent energy density for the electrostatic case turns out to be, $ \dfrac{1}{2} \overrightarrow{E}^2$ and for the case $\overrightarrow{E}$ and $\overrightarrow{D}$ point in the same direction, which is satisfied if $E= |E|> \dfrac{f}{\sqrt{2}}$ , then, $ \dfrac{1}{2} \overrightarrow{E}^2=\dfrac{1}{2} \overrightarrow{D}^2+\dfrac{f}{\sqrt{2}}D+\dfrac{f^2}{4} $, so that it indeed contains a term linear w.r.t. $D= |D|$ as argued by  't Hooft \ct{tHooft-03}. 
The vacuum state is degenerated and is defined by the states with $\overrightarrow{D} = 0$, notice that a charge source generates  $\overrightarrow{D}$, not  $\overrightarrow{E}$, furthermore, the states with  
$\overrightarrow{D} = 0$
are solutions of the equations of motion, not so states with  $\overrightarrow{E} = 0$, in fact at such a point in field space the equations of motion are not even well defined.
However, for all the solutions studied in this paper the excitations "over the vacuum" will satisfy  $E=|\overrightarrow{E}| > \dfrac{f}{\sqrt{2}}$, while  $E= \dfrac{f}{\sqrt{2}}$, will correspond to the "vacuum configuration" of the electrostatic theory, as  it will be discussed. Similar connection between $\overrightarrow{D}$ and $\overrightarrow{E}$ has been considered as an example of a "classical model of confinement" in Ref. \ct{GG-7} and analyzed generalizing the methods developed for the "leading logarithm model" in Ref.  \ct{GG-8}.

The natural appearance of the ``square-root'' Maxwell term in the effective
gauge field action \rf{flatmodel} was further motivated by  't Hooft \ct{tHooft-03}
who has proposed that such gauge field actions are adequate for describing 
confinement (see especially Eq.(5.10) in \ct{tHooft-03}). He has in particular described a
consistent quantum approach in which these kind of terms play the
role of ``infrared counterterms''.
Also, it has been shown in first three refs.\ct{GG-1}--\ct{GG-6} that the square root of 
the Maxwell term naturally arises as a result of spontaneous breakdown of scale symmetry of 
the original scale-invariant Maxwell theory with $f$ appearing as an integration 
constant responsible for the latter spontaneous breakdown.

\smallskip

We will consider the simplest coupling to gravity of the nonlinear gauge field system
with a square-root of the Maxwell term. 
The relevant action is given by (we use units with Newton constant $G_N=1$):
\br
S = \int d^4 x \sqrt{-g} \Bigl\lb \frac{R(g)-2\Lambda}{16\pi} + L(F^2)\Bigr\rb \quad ,\quad
L(F^2) = - \frac{1}{4} F^2 - \frac{f}{2} \sqrt{- F^2} \; ,
\lab{gravity+GG} \\
F^2 \equiv F_{\k\l} F_{\m\n} g^{\k\m} g^{\l\n} \quad ,\quad 
F_{\m\n} = \pa_\m A_\n - \pa_\n A_\m \; .
\nonu
\er
Here $R(g)$ is the scalar curvature of the space-time metric
$g_{\m\n}$ and $g \equiv \det\Vert g_{\m\n}\Vert$, $f$ is a positive coupling constant. 

Let us note that the Lagrangian $L(F^2)$ in \rf{gravity+GG} contains both the 
usual Maxwell term as well as a non-analytic function of $F^2$ and thus it
is a {\em non-standard} form of nonlinear electrodynamics. In this way it is 
significantly different from the original purely ``square root'' Lagrangian 
$- \frac{f}{2}\sqrt{F^2}$ first proposed by Nielsen and Olesen \ct{N-O-1} to
describe string dynamics (see also refs.\ct{N-O-2,N-O-3}). The Nielsen and Olesen action was designed to be applicable only to "magnetic dominated" configuration, for electric dominated configurationd the square root becomes imaginary.
\noindent
In contrast, here will be interested in the "electric" sector of the theory and  notice that now for magnetic dominated configurations the argument inside the square root becomes negative and the square root itself imaginary.  This could be a real effect (that gauge fields must be electrically dominated), or since the action  is not analytic, it is of course possible to construct a simple modification that allows magnetic dominated configurations by taking absolute value inside the square root. This will not affect the electrostatic sector of the theory, but it will be harder to motivate (for example from spontaneous symmetry breaking of scale invariance), but that discussion would go well beyond the purpose of the present paper, which is  to see what can a theory that provides confining theory in flat space do in the presence of horned space times. 

As done in \ct{tHooft-03} we also mean to use Lagrangian \rf{gravity+GG} to describe a truncation of the non Abelian theory, where for simplicity we take the gauge field potential in a specific direction in color space (so commutator terms vanish). Let us also remark that one could start with the non-Abelian version of the 
gauge field action in \rf{gravity+GG}. Since we will be interested in static 
spherically symmetric solutions, the non-Abelian gauge theory effectively reduces 
to an Abelian one as pointed out in the ref.\ct{GG-1}.

The corresponding equations of motion of \rf{gravity+GG} reads:
\be
R_{\m\n} - \h g_{\m\n} R+\Lambda g_{\m\n} = 8\pi T^{(F)}_{\m\n} \; ,
\lab{einstein-eqs}
\ee
where
\be
T^{(F)}_{\m\n} =
L(F^2)\,g_{\m\n} - 4 L^{\pr}(F^2) F_{\m\k} F_{\n\l} g^{\k\l} \; ,
\lab{T-F}
\ee
and
\be
\pa_\n \(\sqrt{-g} L^{\pr}(F^2)  F_{\k\l} g^{\m\k} g^{\n\l}\)=0 \; ,
\lab{GG-eqs}
\ee
where $L^{\pr} (F^2)$ denotes derivative w.r.t. $F^2$ of the function $L(F^2)$ in 
\rf{gravity+GG}. 

A note concerning the vacuum of the theory is in order here. As opposed to the standard Maxwell theory, the vacuum of the theory is not obtained for zero gauge field strength $F_{\m\n}=0$. Instead, the vacuum of the theory in obtained for configuration such that $L^{'}(F^2)=0$, notice that in such case $T^{(F)}_{\m\n} \propto g_{\m\n}$, that is, it gives a "cosmological constant type contribution", also notice that this is obtained for $F^2 \neq 0$, which agrees with the notion that the vacuum of confining theory contains gauge field condensates.

In our preceding paper \ct{grav-cornell} we have shown that the gravity-gauge-field
system \rf{gravity+GG} possesses static spherically symmetric solutions
with a radial electric field containing both Coulomb and {\em constant} vacuum
pieces:
\be
F_{0r} = \frac{\vareps_F f}{\sqrt{2}} + \frac{Q}{\sqrt{4\pi}\, r^2} 
\quad ,\quad \vareps_F = \mathrm{sign}(Q) \; ,
\lab{cornell-sol}
\ee
\noindent
the sign of the second term, determined by $\vareps_F$, which is the field strength $F_{0r}$ divided by its absolute value has the sign of $F_{0r}$ itself and looking at small enough $r$, we see that this sign is determined by $Q$, since there the Coulomb part dominates. We see therefore that both contributions in \ref{cornell-sol} have the same sign and therefore $|F_{0r}|>\frac{f}{\sqrt{2}}$, i.e., bigger than its vacuum value,
and the space-time metric we have:  
\br
ds^2 = - A(r) dt^2 + \frac{dr^2}{A(r)} + r^2 \bigl(d\th^2 + \sin^2 \th d\vp^2\bigr)
\; ,
\lab{spherical-static} \\
A(r) = 1 - \sqrt{8\pi}|Q|f - \frac{2m}{r} + \frac{Q^2}{r^2} - \frac{\Lambda_{eff}}{3} r^2 \; ,
\lab{RN-dS+const-electr}
\er
is Reissner-Nordstr{\"o}m-de-Sitter-type space, where $\Lambda_{eff}=2\pi f^2+\Lambda$

Appearance in \rf{RN-dS+const-electr} of a ``leading'' constant term different from 1
resembles the effect on gravity produced by a spherically symmetric ``hedgehog''
configuration of a nonlinear sigma-model scalar field with $SO(3)$ symmetry, that is the field of a global monopole 
\ct{BV-hedge} (cf. also \ct{Ed-Rab-hedge}).

\section{Generalized Levi-Civita-Bertotti-Robinson Space-Times}
\label{gen-BR}
Here we will look for static solutions of Levi-Civita-Bertotti-Robinson type 
\ct{LC-BR-1}--\ct{LC-BR-3} of the system \rf{einstein-eqs}--\rf{GG-eqs}, this was studied in \ct{hidingLLB},\ct{hide-confine}. Namely, with 
space-time geometry of the form $\cM_2 \times S^2$ where $\cM_2$ is some two-dimensional
manifold:
\be
ds^2 = - A(\eta) dt^2 + \frac{d\eta^2}{A(\eta)} 
+ a^2 \bigl(d\th^2 + \sin^2 \th d\vp^2\bigr) \;\; ,\;\;  
-\infty < \eta <\infty \;\; ,\;\; a = \mathrm{const}
\; ,
\lab{gen-BR-metric}
\ee
and the electromagnetic field is given by
\be
F_{\m\n} = 0 \;\; \mathrm{for}\; \m,\n\neq 0,\eta \quad ,\quad
F_{0\eta} = F_{0\eta} (\eta) \; ;
\lab{electr-static}
\ee

The gauge field equations of motion become:
\be
\pa_\eta \Bigl( F_{0\eta} - \frac{\vareps_F f}{\sqrt{2}}\Bigr) = 0
\quad ,\quad \vareps_F \equiv \mathrm{sign}(F_{0\eta}) \; ,
\lab{GG-eqs-0}
\ee
yielding a constant vacuum electric field:
\be
F_{0\eta} = c_F = \mathrm{arbitrary ~const} \; .
\lab{const-electr}
\ee
The (mixed) components of energy-momentum tensor \rf{T-F} read:
\be
{T^{(F)}}^0_0 = {T^{(F)}}^\eta_\eta = - \h F^2_{0\eta} \quad ,\quad
T^{(F)}_{ij} = g_{ij}\Bigl(\h F^2_{0\eta} - \frac{f}{\sqrt{2}}|F_{0\eta}|\Bigr)
\; .
\lab{T-F-electr}
\ee
Taking into account \rf{T-F-electr}, the Einstein eqs.\rf{einstein-eqs} for
$(ij)$, where $R_{ij}=\frac{1}{a^2} g_{ij}$ because of the $S^2$ factor in
\rf{gen-BR-metric}, yield:
\be
\frac{1}{a^2} = 4\pi c_F^2+\Lambda 
\lab{einstein-ij}
\ee
The $(00)$ Einstein eq.\rf{einstein-eqs} using the expression 
$R^0_0 = - \h \pa_\eta^2 A$ (ref.\ct{Ed-Rab-1}; see also \ct{Ed-Rab-2}) becomes:
\be
\pa_\eta^2 A = 8\pi h(|c_F|) \quad ,\quad  h(|c_F|) \equiv c_F^2-\sqrt{2}f|c_F|-\frac{\Lambda}{4\pi}
\lab{einstein-00}
\ee
In the particular case where $\Lambda=0$, studied in \ct{hidingLLB}, we arrive at the following three distinct types of
Levi-Civita-Bertotti-Robinson solutions for gravity coupled to the
non-Maxwell gauge field system \rf{gravity+GG}:

(i) $AdS_2 \times S^2$ with strong constant vacuum electric field
$|F_{0\eta}| = |c_F|>\sqrt{2}f$, where $AdS_2$ is two-dimensional anti-de Sitter 
space with:
\be
A(\eta) = 1+ 4\pi |c_F| \(|c_F| - \sqrt{2}f\)\,\eta^2
\lab{AdS2}
\ee
in the metric \rf{gen-BR-metric}.

(ii) $M_2 \times S^2$ with constant vacuum electric field 
$|F_{0\eta}| = |c_F|=\sqrt{2}f$, where $M_2$ is the flat two-dimensional 
 space with:
\be
A(\eta) = 1 
\lab{Rindler2}
\ee
in the metric \rf{gen-BR-metric}.

(iii)  $dS_2 \times S^2$ with weak constant vacuum electric field
$|F_{0\eta}| = |c_F|<\sqrt{2}f$, where $dS_2$ is two-dimensional de Sitter space with:
\be
A(\eta) = 1 - 4\pi |c_F| \(\sqrt{2}f - |c_F|\)\,\eta^2
\lab{dS2}
\ee
 For the special value $|c_F| = \frac{f}{\sqrt{2}}$
we recover the Nariai solution \ct{nariai-1,nariai-2} with $A(\eta) = 1 - 2\pi f^2 \eta^2$ 
and equality (up to signs) among energy density, radial and transverse pressures:
$\rho = - p_r = - p_{\perp} = \frac{f^2}{4}$ (${T^{(F)}}^\m_\n = \mathrm{diag} \(-\rho,p_r,p_{\perp},p_{\perp}\)$).

In all three cases above the size of the $S^2$ factor is given by 
\rf{einstein-ij}. Solutions \rf{Rindler2} and \rf{dS2} are new ones and are specifically due to 
the presence of the non-Maxwell square-root term in the gauge field Lagrangian \rf{gravity+GG}.

In this paper we will consider the case $\Lambda\neq 0$ and demand that $\Lambda_{eff}=0$. This leaves us only with solutions similar to \rf{AdS2} although with a different dependence on $|c_F|$ (see eq. \rf{AntiDeSitter})

\section{Matching through a regular (time-like) thin shell }

Now, we want to discuss the matching of \rf{spherical-static} to \rf{gen-BR-metric} at a spherically symmetric wall. The metric induced at the wall has to be well defined, and the coefficient of purely angular displacements $d\Omega^2$ in \rf{spherical-static} and \rf{gen-BR-metric} has to agree at the position of the wall, to give the same value of $ds^2$. Therefore, at the wall 

\be
 r=a
 \lab{location}
\ee

The equations of motion of a thin layer in GR have been obtained by W.Israel \cite{Israel-66}, we now briefly review these results. To get those equations it is useful to define a Gaussian Normal Coordinate system in a neighbourhood of the wall as follows; denoting the $2+1$ dimensional hypersurface $\Sigma$ and introducing a coordinate system on $\Sigma$, two are taken to be the angular variables $\theta,\phi$, which are always well defined up to an overall rotation for a spherically symmetric configuration. For the other coordinate in the wall, one can use the proper time variable $\tau$ that would be measured by an observer moving along with the wall. The fourth coordinate $\xi$ is taken as the proper distance along the geodesics intersecting $\Sigma$ orthogonally. We adopt the convention that $\xi$ is taken to be positive in the Reissner-Nordstr{\"o}m-de-Sitter-type regime and negative in the Generalized Levi-Civita-Bertotti-Robinson regime, and $\xi=0$ is of course the position of the wall. Thus the full set of coordinates is given by $x^\mu=(\tau,\theta,\phi,\xi)$,   
in this coordinates 
\be
g^{\xi\xi}=g_{\xi\xi}=1 \quad , \quad g^{\xi i}=g_{\xi i}=0 
\ee

Also, we define $n_{\mu}$ to be the normal to an $\xi=constant$ hypersurface, which in Gaussian normal coordinates has the simple form $n_{\mu}=n^{\mu}=(0,0,0,1)$. We then define the extrinsic curvature corresponding to each $\xi=constant$ hypersurface, which is a $3$-dimensional tensor whose components are defined by
\be
K_{ij}=n_{i;j}=\frac{\pa n_i}{\pa x^j}-\G^{\alpha}_{ij}n_{\alpha}=-\G^{\xi}_{ij}=\frac{1}{2}\pa_{\xi}g_{ij} 
\lab{extrinsic curvature}
\ee
As we can see, the extrinsic curvature gives the change of the metric in the direction perpendicular to the surface.

In terms of these variables, the Einstein's equations take the form;\\
\begin{equation}
 \begin{array}{l l}
  G^{\xi}_{\xi}\equiv -\frac{1}{2}{^{(3)}R}+\frac{1}{2}\left[(Tr K)^2-Tr(K^2) \right]=8\pi G T^{\xi}_{\xi} & \\ \\
  G^{\xi}_{i}\equiv K^{m}_{i|m}-(TrK)_{|i}=8\pi G T^{\xi}_{i}
  \end{array} 
  \lab{first Einstein FQ}
\end{equation}
and 
\begin{equation}
 \begin{array}{l l}
G^{i}_{j}\equiv {^{(3)}G^{i}_{j}}-\left(K^{i}_{j}-\delta^{i}_{j} TrK\right)_{,\xi} -\left(Tr K\right)K^{i}_{j} &\\ \\ \qquad +\frac{1}{2}\delta^{i}_{j} \left[(Tr K)^2+Tr(K^2) \right]=8\pi G T^{i}_{j}
 \end{array} 
  \lab{second Einstein FQ}
\end{equation}

where the subscript vertical bar denotes the $3$-dimensional covariant derivative in the $2+1$ dimensional space of coordinates $(\tau,\theta,\phi)$, and comma denotes an ordinary derivative. Also quantities like $^{(3)}R$, $^{(3)}G^{i}_{j}$, etc. are to be evaluated as if they concerned to a purely $3$-dimensional metric $g_{ij}$, without any reference as to how it is embedded in the higher four dimensional space. 

By definition, for a thin wall, the energy-momentum tensor $T^{\mu\nu}$ has a delta function singularity at the wall, so one can define a surface stress energy tensor $S^{\mu\nu}$
\be
T^{\mu\nu}=S^{\mu\nu}\delta(\xi)+regular\space\ terms
\lab{Energy momentum tensor}
\ee
When the energy momentum tensor \rf{Energy momentum tensor} is inserted into the field equations \rf{first Einstein FQ},\rf{second Einstein FQ}, we obtain that \rf{first Einstein FQ} are satisfied automatically, provided that they are satisfied for $\xi\neq0$ and provided that $g_{ij}$ is continuous at $\xi=0$ (so that $K_{ij}$ does not acquire a $\delta$-function singularity). Eq.\rf{second Einstein FQ} however, when integrated from $\xi=-\varepsilon$ to $\xi=\varepsilon$ ($\varepsilon\rightarrow 0$ and $\varepsilon >0$), leads to the discontinuity condition 
\be
S^i_j=-\frac{1}{8\pi G}\left[ \gamma^i_j-\delta^i_j Tr\gamma \right]
\ee
or equivalently
\be
\gamma^i_j=-8\pi G \left[ S^i_j-\frac{1}{2}\delta^i_j Tr S \right]
\lab{JCE}
\ee
where
\be
 \gamma_{ij}=\lim_{\epsilon\rightarrow0}\left[K_{ij}(\xi=+\epsilon)-K_{ij}(\xi=-\epsilon) \right]
\ee
is therefore the \emph{jump} of extrinsic curvature across $\Sigma$.

Local conservation of $T_{\mu\nu}$ and the demand of spherical symmetry gives a surface stress-energy tensor of the form
\be
S^{\mu\nu}=\sigma(\tau)U^{\mu}U^{\nu}-\omega(\tau)[h^{\mu\nu}+U^{\mu}U^{\nu}]
\lab{surface energy tensor}
\ee
where $h^{\mu\nu}=g^{\mu\nu}-n^{\mu}n^{\nu}$ is the metric projected onto the hypersurface of the wall, and $U^{\mu}=(1,0,0,0)$ is the four velocity of the wall. 

In \rf{surface energy tensor}, $\sigma$ has the interpretation of energy per unit surface, as detected by an observer at rest with respect to the wall, and $\omega$ has the interpretation of surface tension. For a given equation of state 
$p=p(\sigma)$ ($p=-\omega$) the local energy momentum at the wall gives $d(r^2\sigma)=-pd(r^2)$,
but since we have $r=Const$ at the matching point, we get $\sigma=Const$ and from the generic equation of state $p=p(\sigma)=Const$. For the shell at the junction $r=a$, the angular coordinates $( \theta , \phi)$ must be identified. Also the radial coordinate of the shell, as seen from the outside, is clearly $r=a$, but this smooth sewing so that the requirement that the induced metrics on the shell from the inside and outside giving the same result on the shell, and therefore being well defined, is not enough to determine $ \eta $ in \rf{gen-BR-metric}, which will have a non trivial time dependence $\eta=\eta(t)$ to be determined by the Israel junction conditions.   

In addition to this, the timelike brane can also have a delta function charge densnity $j^\mu=\delta ^{\mu}_0 q \delta(\xi)$ coming from the discontinuity in the gauge field strength across the matching at $r=a$. 

\be
[F_{0\nu}]_{r=a}=q
\ee  
which can also be defined in terms of the electric flux, or more clearly, by defining the electric displacement field $D_{0\nu}$ in the two regions which in the present case is significantly different from the electric field $F_{0\nu}$ due to the presence of the "square-root" Maxwell term
\be
 D_{0r}|_{r=a}-D_{0\eta}=q 
\lab{gauge field discontinuity}
\ee
where $D_{0r}=\left(1-\frac{f}{\sqrt{2}|F_{0r}|}\right)F_{0r}$ and $D_{0\eta}=\left(1-\frac{f}{\sqrt{2}|F_{0\eta}|}\right)F_{0\eta}$

\bigskip

We now consider the matching through a thin spherical wall of the Reissner-Nordstr{\"o}m-de-Sitter-type regime (denoted by "+") to the Generalized Levi-Civita-Bertotti-Robinson regime (denoted by "-"). First consider the discontinuity of $K_{\theta\theta}$. Using \rf{location},\rf{JCE} and \rf{surface energy tensor} we get 
\be
 K_{\theta \theta}^-- K_{\theta \theta}^+=4\pi \sigma a^2
\ee

for the Levi-Civita-Bertotti-Robinson space $g_{\theta\theta}=a^2=constant$, so that $K_{\theta \theta}^-=0$ according to \rf{extrinsic curvature}. Therefore $ K_{\theta \theta}^+=-4\pi \sigma a^2$
\be
 \sqrt{A_{0}(a)}=-4\pi \sigma a
 \lab{theta component}
\ee
where $-A_o$ denotes the $0-0$ component of the metric "outside", on the $r>a$ region.
So, if we are dealing with a standard static space time outside (for example in the case of Schwarzschild space using only region I), the above equation implies $\sigma<0$. That is, the matching of the tube space time with the "normal" outside space will require negative energy densities. We could insist on $\sigma>0$ but this requires the use of all regions of Kruskal space \ct{eduardo-wh}(this allows the square root in \rf{theta component} to be negative \ct{eduardo-wh}) and therefore the resulting wormhole is not traversable, we will not follow this approach in this paper.
\smallskip

The $\phi\phi$ component of equation \rf{JCE} gives the same information due to the spherical symmetry of the problem. The additional information will come from the $\tau\tau$ component, which reads   
\be
  K_{\tau \tau}^-- K_{\tau \tau}^+=4\pi (\sigma -2 \omega)
  \lab{tau component}
\ee
from $U^\mu n_\mu=0$ we have $U^\mu n_{\mu ; \nu}=-n_{\mu}U^{\mu}_{;\nu}$, and the $K_{\tau \tau}$ component then takes the form
\be
K_{\tau \tau}=-n_{\mu}\left(\frac{d U^\mu}{d\tau}+\Gamma^{\mu}_{\alpha\beta}U^{\alpha}U^{\beta}\right)
\lab{K tau tau}
\ee

Using this expression for the $\tau\tau$ component of the extrinsic curvature, 
its discontinuity equation gives then,
 
\be
 -\frac{1}{\dot{\eta}} \frac{d}{d \tau}\left(\sqrt{A_i(\eta)+\dot{\eta}^2} \right)+\frac{1}{2\sqrt{A_0(a)}}\frac{\partial A_o(r)}{\partial r} \v_{r=a}=4\pi (\sigma -2 \omega)
\ee
where $-A_i(\eta)$ denotes the $0-0$ component of the metric "inside", on the tube like compactified region.

Using \rf{theta component}
\be
-\frac{1}{\dot{\eta}} \frac{d}{d \tau}\left(\sqrt{A_i(\eta)+\dot{\eta}^2} \right)-\frac{1}{8\pi \sigma a}\frac{\partial A_o(r)}{\partial r} \v_{r=a}=4\pi (\sigma -2 \omega)
\lab{equation for eta}
\ee
multiplying by $\dot{\eta}$ and defining
\be
\Delta \equiv \frac{1}{8\pi \sigma a}\frac{\partial A_o(r)}{\partial r} \v_{r=a}+4\pi (\sigma -2 \omega) 
\ee
we can see that equation \rf{equation for eta} becomes a total derivative of the proper time $\tau$ 
\be
\frac{d}{d \tau}\left[\sqrt{A_i(\eta)+\dot{\eta}^2}+\Delta \eta \right]=0
\ee
which, upon integrating, we obtain
\be
\sqrt{A_i(\eta)+\dot{\eta}^2}+\Delta \eta=E=const
\ee
taking $\Delta$ to the right hand side and then squaring, we finally obtain
\be
\dot{\eta}^2=(\Delta \eta-E)^2-A_i(\eta)
\lab{eta potential}
\ee
where the general expression of $\Delta$, using the explicit form \rf{RN-dS+const-electr} for the uncompactified region, is 
\be
\Delta \equiv \frac{1}{4\pi \sigma a^4}(ma -Q^2-\Lambda_{eff}a^4)+4\pi (\sigma -2 \omega) 
\lab{diff of delta}
\ee

\smallskip

\section{Asymptotically flat, finite energy solutions, imply hiding the electric flux}

Let us assume our ground state is just flat space, and on top of that we would like to build finite energy excitations. In the first place this means that $\Lambda_{eff}=0$, but still, this is not enough to ensure asymptotic flatness of finite energy. Indeed, if $Q \neq 0$, the leading behaviour of the metric is not flat, but rather "hedhehog type" ,
\ct{BV-hedge}, \ct{Ed-Rab-hedge}, which have energy momentum tensor that decrease only as the square of the radius for large distances, which of course are infinite energy solutions.

This is of course consistent with the notion that in a confining theory an isolated charge  has an associated infinite energy. Here, if we add the horned particle to the problem, the isolated charge can have finite energy, provided it sends all the electric flux it produces to the tube region, this requires the vanishing of the external Coulomb part of the electric field, or $Q=0$.

When $Q=0$, then, according to \rf{cornell-sol}, the electric field outside has magnitude $|F_{0r}|=\frac{f}{\sqrt{2}}$. Therefore, in the $r>a$ region, the displacement field is equal to zero $D_{0r}=0$. Thus, the absence of Coulomb field, our choice $\Lambda_{eff}=0$, and assuming absence of magnetic field, the outer region then simply becomes a Schwarzschild solution with vacuum electric field, which has constant magnitude $\frac{f}{\sqrt{2}}$ . But for this value of the field strength we have that the gauge field Lagrangian has a minimum, in fact, $L^{\pr}(F^2)=0$, so using only that $F^2=f^2$ the equation of motion for the gauge field is satisfied automatically, that is to say, we now do not need the electric field to be radial, the orientation is now completely arbitrary, once $F^2=f^2$ is satisfied. In this disordered vacuum, where the electric field with constant magnitude does not point in one fixed direction, a test charged particle will not be able to get energy from the electric field, instead, it will undergo a kind of Brownian motion, therefore \emph{no} Schwinger pair-creation mechanism will take place.      

\bigskip

From \rf{diff of delta} and the above discussion, $\Delta$ takes the following simple form
\be
\Delta\rightarrow \frac{m}{4\pi \sigma a^3}+4\pi (\sigma -2 \omega)
\ee
and from \rf{einstein-00}
\be
 \partial_\eta^2 A_i(\eta)=8\pi D(|c_F|)^2  
\ee
where $D(|c_F|)$ is the displacement field in the compactified region, this leads to a metric of the anti de Sitter form for two dimensional space factor,
\be
A_i(\eta)=1+4\pi D^2\eta^2
\lab{AntiDeSitter}
\ee
equation \rf{eta potential} then becomes 
\be
 \dot{\eta}^2=E^2-1+(\Delta^2-4\pi D^2)\eta^2-2\Delta E\eta
\ee
which can be cast onto the form 
\be
 \dot{\eta}^2+(4\pi D^2-\Delta^2)\left(\eta+\frac{\Delta E}{4\pi D^2-\Delta^2}\right)^2=\frac{4\pi D^2 E^2}{4\pi D^2-\Delta^2}-1
\ee

\begin{figure}[H]
  \centering
  \subfloat[stable]{\lab{fig:stable}\includegraphics[width=0.5\textwidth]{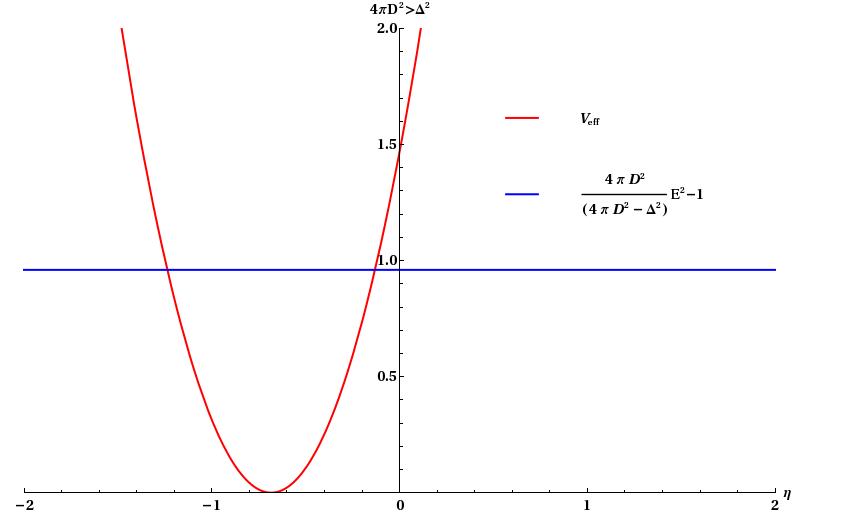}}  
  \subfloat[unstable]{\lab{fig:unstable}\includegraphics[width=0.5\textwidth]{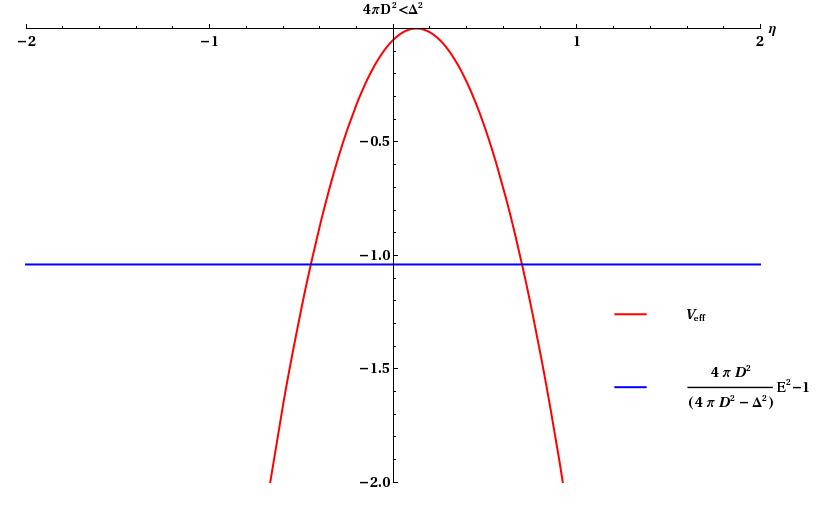}}
  \caption{An illustration of the possible solutions; in the first one (a), we have a stable solution with the "Energy" bounded $E^2>1-\left(\frac{\Delta}{\sqrt{4\pi}D}\right)^2\equiv E^2_{min}$. The second case (b), present an unstable solution. In this case there is no bound on the "Energy" since $\frac{4\pi D^2 E^2}{4\pi D^2-\Delta^2}-1<0$ for all values of $E$  }
  \lab{potential}
\end{figure}

Going back to Eq.\rf{gauge field discontinuity}, the charge of the time like shell, call it $K$, will be determined by the flux produced only by the tube region $D(|c_F|)$. Using $D_{0r}=0$, we have $D_{0\eta}=-q=-K/4\pi a^2$. Notice that $D$ and therefore $q$ must be different from zero since Eq.\rf{einstein-ij} together with $\Lambda_{eff}=0$  imply $|c_F|>\frac{f}{\sqrt{2}}$, otherwise the the radii $a$ will be ill-defined or infinite. 

\smallskip

Using now the condition $\Lambda_{eff}=0$, we obtain from \rf{einstein-ij}

\be
\frac{1}{a^2} = 4\pi \left(|c_F| - \frac{f}{\sqrt{2}}\right)\left(|c_F| + \frac{f}{\sqrt{2}}\right)    
\lab{newa}
\ee

Expressing $|c_F|$ in terms of $D$ and then $D$ in terms of $|K|$, and taking into account the Sign of $c_F$ and $q$ we see that \rf{newa} provides a quadratic equation for the charge $|K|$ in terms of $a$ and $f$
\be
0=\frac{|K|^2}{4\pi a^2}+\sqrt{2}f|K|-1
\ee
which has a positive solution for all possible values of those parameters, since the discriminant of such quadratic equation is manifestly 
positive.
\begin{figure}[H]
  \centering
 {\lab{fig:charge}\includegraphics[width=0.5\textwidth]{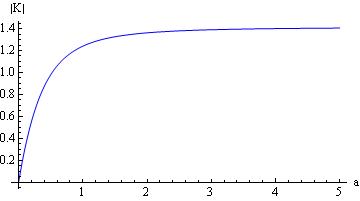}}         
  \caption{The charge of the time like shell at the throat of the wormhole as a function of the radii $a$.}
  \lab{charge}
\end{figure}
This completes the proof that there are indeed consistent, horned particle solutions where all the electric flux of the charged particle at the throat flows into the tube region, furthermore, these are the only finite energy solutions.

\section{Discussion and Perspectives for future Research}

The charge-hiding effect by a wormhole, which was studied for the case where gravity/gauge-field system is self-consistently interacting with a charged lightlike brane (LLB)as a matter source, is now studied for the case of a time like brane. From the demand that no surfaces of infinite coordinate time redshift appear in the problem we are lead now to a horned like particle where the horn region of the particle is completely accessible from the outside region containing the $r\rightarrow\infty$ region and vice versa, according to not only the traveller that goes through the shell (as was the case for the LLB), but also to a static external observer. This requires negative surface energy density for the shell sitting at the throat. 
We study a gauge field subsystem which is of a special non-linear form containing a square-root of the 
Maxwell term and which previously has been shown to produce a QCD-like confining gauge field dynamics in flat space-time. 
The condition of finite energy of the system or asymptotic flatness on one side of the horned particle implies that the charged object sitting at the throat expels all the flux it produces into the other side of the horned particle, which turns out to be of a ``tube-like'' nature. An outside  observer in the asymptotically flat universe detects, therefore, apparently neutral object. The hiding of the electric flux in the horn region is the only possible way that a truly charged particle can still be of finite energy, which points to the physical relevance of such solutions, even though there is the need of negative energy density at the throat, which can be of quantum mechanical origin.

In addition to the "hiding" effect, one can also study the "confinement" \ct{hide-confine}, where instead of considering just the case of a matching of an external uncompactified region with a compactified tube region, we consider two asymptotically flat regions connected by a tube region and at the two points where we match the corresponding uncompactified to the tube region we have a brane, the system will contain a brane plus associated "antibrane" at the other matching point. In the case where Light like Branes are used at the matching points, the branes are, classically at least, located at fixed coordinate locations \ct{hide-confine}, for time like branes, this will not be generically the case, so the brane could in principle collide with its antibrane. These possibilities will be studied in a future publication.

We have seen, for zero $Q$, that in the exterior region any configuration satisfying ${\mathcal L^{\prime}(F^{2})} = 0$, or (or $D = 0$ in the electrostatic case) is a solution of the gauge field equation,  obtaining in the vacuum region a "disordered ferroelectric state". This degeneracy of the outside state is good from point of view the high entropy content of the configuration, since it means that there is a great many ways, infinite indeed, to achieve this matching with zero Coulomb field outside. 
This subject and its presumably favorable consequences for the stability of this vacuum state with gauge field condensation will be studied in future publications.

Going back to the hiding effect studied in this paper, it is interesting to consider now whether these solutions where the charged particle expels the flux exclusively in the direction of the tube region can take place in nature, or whether this is just a mathematical exercise. This question naturally relates to whether the negative energy density at the throat is physically realizable. To start with, we know that negative energy densities can be achieved through quatum correctios, as it has been discussed in \ct{FordRoman}, \ct{RomanFord}. These authors  have however found, studying some field theory models that to build up regions of negative energy density, one must "pay" by building compensating regions with positive energy density elsewhere. More generically this means that one cannot arbitrarily assign some negative energy density to some region of space and just blame quantum fluctuations for that. For explicit calculations showing that quantum fluctuations can be the origin of negative energy densities see \ct{Negative energy},\ct{Static negative energies}.

It i interesting that the gauge field model that produces confinement may give the possibility of obtaining negative energy densities. In the solutions studied so far in this paper, this has not been the case because the electic fields involved in all of the solutions considered in this paper are stronger that the vacuum value, recall for example that in the tube region $|c_F|>\frac{f}{\sqrt{2}}$. If we look at lower field strengths absolute values lower than the vacuum value, if they could somehow be obtained (may be as a result of quantum fluctuations), then negative energy densities could result. To see this, just consider the flat space situation and then recall that for static field configurations, the model \rf{flatmodel} yields the following electric displacement field $\overrightarrow{D} = \overrightarrow{E} - \dfrac{f}{\sqrt{2}}\dfrac{\overrightarrow{E}}{|\overrightarrow{E}|}$. The pertinent energy density for the electrostatic case turns out to be, $ \dfrac{1}{2} \overrightarrow{E}^2$ and for the case $\overrightarrow{E}$ and $\overrightarrow{D}$ point in opposite directions, which is satisfied if $E= |E|< \dfrac{f}{\sqrt{2}}$ , then, $ \dfrac{1}{2} \overrightarrow{E}^2=\dfrac{1}{2} \overrightarrow{D}^2- \dfrac{f}{\sqrt{2}}D+\dfrac{f^2}{4} $, so that the term linear w.r.t. $D= |D|$ is negative now. 
For low values of $D$, this dominates over the quadratic contribution and therefore, we get an energy density lower than that of the vacuum state, which is the state with  $E=|E|= \dfrac{f}{\sqrt{2}}$ (or $D = 0$). With the appropriate bare cosmological constant chosen here, the vacuum state vacuum energy density is zero, so the low electric field configurations produce then a negative vacuum energy density. Whether this way of achieving negative energy densities can be used to achieve horned particle hiding charge remains an interesting subject for future research.

In the case of the charge hiding in a horned particle studied here, we have the special situation where the hiding of the electric flux behind the horn region, and no flux of $D$ going in the outside region is the only possible way that a truly charged particle (in a model with confining dynamics) can still be of finite energy, while still remaining truly charged, which points to the physical relevance of such solutions, even though there is the need of negative energy density at the throat, which can be of quantum mechanical origin or may be just classical (by means of low electric field strengths). One can then argue that a variational approach to the problem, based on minimization of energy must produce indeed select the 
appropriate state that gives rise  the necessary negative energy density at the throat of the horned particle, so that the flux lines can be now redirected into the horn region and make the finite energy solution possible.


\section*{Acknowledgments}
We would like to thank  A. Kaganovich, E. Nissimov, S. Pacheva for very usefull conversations.


\end{document}